
\magnification=\magstep1  \overfullrule=0pt
\advance\hoffset by 0.65truecm
\def\lb{\lbrack}\def\rb{\rbrack}  \def\q#1{$\lb#1\rb$}
\def\pl{{\rm Phys.\ Lett.\ B\ }} \def\np{{\rm Nucl.\ Phys.\ B\ }}
\def\CMP{{\rm Commun.\ Math.\ Phys.\ }}
\def\intj{{\rm Int.\ J.\ Mod.\ Phys.\ A\ }}
\def\bn{\bigskip\noindent} \def\mn{\medskip\smallskip\noindent}
\def\sn{\smallskip\noindent} \def\crns{\cr\noalign{\smallskip}}
\def\l{\hbox{$\cal L$}}  \def\a{\hbox{$\cal A$}}
 
\def\irhwm{\hbox{irreducible highest weight mo\-dule}}
\def\ie{i.e.\ } 
 
\def\g{\hbox{$\cal G$}} 
\def\P{\hbox{$\cal P$}}   
\def\Z{\hbox{{\rm Z{\hbox to 3pt{\hss\rm Z}}}}}
\def\zp{\hbox{${\rm Z{\hbox to 3pt{\hss\rm Z}}}_+$}}
 \font\extras=cmss10 scaled 750
\font\Lar=cmb10 scaled \magstep3  \font\lar=cmr10 scaled \magstep2
\setbox2=\hbox{{{\extras Z}}} \setbox3=\hbox{{{\extras z}}}

\def\C{{\rm C \kern-5.5pt I \ }}
\def\R{\hbox{{\rm I{\hbox to 5.5pt{\hss\rm R}}}}}
\def\menge{\hbox{$\lbrace0,1,\ldots,N\rbrace$}}
\def\mengeK{\hbox{$\lbrace0,1,\ldots,K\rbrace$}}
\def\folge{(a_0,a_1,\ldots,a_M)} 
  
\def\oddots{\mathinner{\mkern1mu\raise1pt\vbox{\kern7pt\hbox{.}}
1 \mkern2mu\raise4pt\hbox{.}\mkern2mu\raise7pt\hbox{.}\mkern1mu}}
\nopagenumbers\rightline{Bonn preprint BONN-HE-92-22, July 1992
\phantom{xxxx}}
\vskip4truecm\centerline{\Lar Virasoro Representations on Fusion Graphs}
\bn\bn \centerline{by} \bn\bn
\centerline {{\lar Johannes Kellendonk} and {\lar Andreas Recknagel}}
\bn\bn\bn \centerline{Universit\"at Bonn, Physikalisches Institut}
\centerline{Nu\ss allee12, D-5300 Bonn 1} \bn\bn\bn
\vskip3truecm
\centerline{\bf Abstract} \bn
  For any non-unitary minimal model with central charge
  $c(2,q)$ the path spaces associated to a certain fusion graph
  are isomorphic to the irreducible Virasoro highest weight modules.
\vfill \leftline{e-mail: UNP02A@IBM.RHRZ.UNI-BONN.DE}
\leftline{\phantom{e-mail: }UNP01A@IBM.RHRZ.UNI-BONN.DE}  \eject
\pageno=1\noindent {\bf 1.}\phantom{xx} During the recent progress in
both conformal field theory \q1 and solvable lattice models in two
dimensions \q2 it has become clear that there are deep connections
between the two areas, which go far beyond the common conviction that
CFT describe universality classes of
critical behaviour of statistical models in
the continuum limit. Apart from the fact that in both branches similar
mathematical structures show up, it has in particular been realized
that the Virasoro algebra plays an important role for lattice models
even away from the critical point and the continuum. E.g., in the
ABF models \q3 and certain generalizations (see \q4 and references
therein) it has been found that in the corner transfer matrix formalism
 the local height probabilities and 1-dimensional configuration sums
can be expressed in terms of Virasoro (and Kac-Moody) characters.
Other generalizations of RSOS models \q5 are believed to yield all the
modular invariant partition functions of Virasoro minimal models
given in \q6. All this indicates that there should exist some
"lattice representations" of the Virasoro algebra, as well as a
direct link between Virasoro and Temperley-Lieb-Jones
algebra, which governs exactly solvable lattice models. Unfortunately,
except for the Ising model \q7, rather little progress has been made
in establishing the explicit connection. The difficulties may partly
originate in the rather complicated energy grading on the path spaces
chosen in \q{3} (see the remarks in section 3) so that it would be
desirable to have different (more natural) path representations
of the Virasoro algebra as well.  \sn
In this letter we will show that, for the simple class of non-unitary
Virasoro minimal models with central charge
$$ c(2,q) = - {(3q-4)(q-3)\over q}\,,  \eqno(1) $$
$q\geq5$ odd, the (degenerate) \irhwm s can be realized as spaces of
paths on a finite graph in a way alternative to the one in \q3,
in particular without any "ground state problems" (see below).
The surprising fact is that the relevant graph occurs as fusion graph
of a distinguished primary field in the $c(2,q)$ minimal model under
investigation.  \sn
In the next section this statement will be made precise, and proved,
whereas section 3 contains some further comments. \bn\bn
\noindent {\bf 2.}\phantom{xx} Starting point of our considerations
are the results of Feigin et al.\ \q{8}, who gave explicit bases of
the (degenerate) \irhwm s ${\cal L}_j$ of the $c(2,q)$ models (see
also $\q{9}\,$). These have highest weights
$$ h_j = {-j(q-2-j)\over2q}\,,\ \ \ j=0,1,\ldots,K   \eqno(2)$$
with $q=2K+3$. Studying the (coarse) structure of the null fields
of the models, and by means of some combinatorial arguments, Feigin
et al.\ could show that a basis of ${\cal L}_j$ is formed by those
elements
$$ L_{m_1}L_{m_2}\cdots L_{m_n}|h_j\rangle  \eqno(3)   $$
with $m_1\geq m_2 \geq \ldots \geq m_n \geq 1$ which furthermore obey
the so-called "difference two condition"
$$ m_i - m_{i+K} \geq 2      \eqno(4) $$
for $1\leq i \leq n-K$, as well as the "initial condition"
$$ \#\lbrace m_i=1\rbrace \leq j\,, \eqno(5) $$
\ie at most the last $j$ of the $m_i$ equal 1. As in \q{8}, we
denote the set of all finite tupels $(m_1,\ldots,m_n)$ of positive
integers satisfying (4,5) by $C^{K,j}$. \mn
In the following we will show that the basis (3-5)
of the irreducible Virasoro modules occuring in the $c(2,q)$ models
can be realized as paths on the (suitably labeled) fusion graph
belonging to $\phi_K$, the primary field with
the lowest conformal dimension showing up in (2).  \sn
The fusion graph $\g_i$ of a primary field $\phi_i$ is the graph
whose connectivity matrix is given by the fusion matrix $N_i$,
\ie it contains one node for each primary field in the QFT ($K+1$
in our cases) and $(N_i)_j^k$ edges (here:\ unoriented) between
the nodes $j$ and $k$, where
$$\phi_i \times \phi_j = \sum_k N_{ij}^k\,\phi_k   \eqno(6) $$
are the fusion rules. (In particular, in any fusion graph $\g_i$
there is one node -- representing the identity field -- joined
to exactly one further node -- representing $\phi_i$ -- by one edge.)
Figure 1 shows some of the fusion graphs relevant for us. \mn
To each graph \g\ we may construct Hilbert spaces $\P(\g,i)$ with
basis vectors given by paths on \g\ starting at node $i$
(running from node to node along the edges). Suppose \g\ is "simply
laced" (at most one edge between two nodes) and that the nodes $j$
carry pairwise distinct integer labels $l(j) \in \menge\,$.
Then each path can equivalently be written as a sequence
$(a_i)_{i \geq 0}$, $a_i \in\menge\,$, and furthermore a sequence
which finally stabilizes at $a_i=0$ for $i>>1$ uniquely represents
an element of the Virasoro Verma module with highest weight
$|h(a_0)\rangle$ by virtue of
$$\folge \mapsto L^{a_M}_M \dots L^{a_1}_1 |h(a_0)\rangle\,.\eqno(7)$$
(we have suppressed the zeroes at the end of the sequence).
This suggests to introduce an energy grading or $L_0$-action
on $\P(\g,i)$ according to
$$L_0 \folge=\bigl(h(a_0)+\sum_{i=0}^M ia_i\bigr)\folge\,,\eqno(8)$$
where $h(a_0)$ is a constant depending on the starting condition
on the paths in $\P(\g,i)$, which is fixed by choosing $a_0$. \mn
On the other hand, equation (7) of course also allows for expressing
the basis elements labeled by tupels in $C^{K,j}$ from above in terms
of "stabilizing" sequences $(a_i)_{i\geq 0}$. It is easy to verify
that the "difference two condition" worked out in \q{8} translates into
$$ a_i + a_{i+1} \leq K\ \  {\rm for\ all}\ i\geq0\,,  \eqno(4') $$
implying in particular $a_i \in \mengeK\,$, whereas the "initial
condition" simply reads  $$ a_1 \leq j \,.  \eqno(5') $$  \bn From
 all this we conclude that the basis (3-5) of
the $c(2,q)$ irreducible representations can indeed be realized
in terms of paths on a labeled graph, provided the latter encodes
the conditions (4') and (5') properly: The graph has to have $K+1$
vertices carrying integer labels from 0 to $K$, and due to (4')
two (not necessarily distinct) nodes are joined by an edge
iff their labels sum up to at most $K$. Condition (5') then just
requires restriction to paths starting at the node labeled $a_0=K-j$.
Stated differently, the graph should have the $(K+1)\times (K+1)$
connectivity matrix
$$ \sigma_K  =  \pmatrix{1 & \cdots &   & 1 & 1    \cr
1 & \cdots & 1 & 1 & 0 \cr 1 & \cdots & 1 & 0 & 0   \cr
\vdots &\oddots& & & \vdots\cr 1 & 0 & \cdots & & 0 \cr}\eqno(9)$$
if the rows and columns are ordered by the label. Note, by the way,
that this labeling is quite a natural one: $l(j) =$ total number of
nodes $-$ number of edges leaving node $j$.
\sn   Thus we have already derived the first part of  \mn
{\bf Proposition:} Fix $q=2K+3$, and let $\g_K$\ be the fusion
graph of the field $\phi_K$ in the $c(2,q)$ minimal model. For $j=
0,1,\ldots,K\,$, label the nodes $j$ (\ie the primary fields $\phi_j$)
by $l(j)=K-j$. Then there is a ($L_0$-grade preserving) bijection
between the \irhwm s ${\cal L}_i$ and the space $\P(\g_K,i)$ of paths
on the labeled graph starting at node $i$ (at label $K-i$). \mn
It remains to identify the connectivity matrix $\sigma_K$ in (9)
with the fusion matrix $N_K$ of the field $\phi_K$ (up to reordering
the rows and columns). This can be done in a straightforward way
using the fusion rules of the $c(2,q)$ minimal models \q1
\def\min{min\lbrace r+s-1,2q-1-r-s\rbrace}
$$ \phi_{(r,1)} \times \phi_{(s,1)} = \sum_{t=|r-s|+1}^{\min}
\phi_{(t,1)}     \eqno(10) $$
for $1\leq r,s \leq q-1,\ r,s,t$ odd. Note that the indices $j$ of the
irreducible highest weight modules
in (2) are related to the more familiar conformal grid
notation in (10) by
$$ j = min\lbrace r-1, q-r -1\rbrace\,. \eqno(11) $$
However, the need to distinguish different subcases arising from
the $min$ in (10) makes the procedure somewhat lengthy. Therefore, we
would rather take advantage of some
results on so-called polynomial fusion rule algebras established in
\q{10}; these are fusion rings possessing a field $\phi$ such
that all other generators are given by polynomials in $\phi$:
$\phi_i = p_i(\phi)$. The $c(2,q)$ models, in particular, are of
this type, and taking $\phi=\phi_1$ (\ie $\phi_{(q-2,1)}$ in the
usual notation), the $p_i$ are
the \v Cebyshev polynomials of the second kind \q{10}:
$$ \eqalign{p_0(x)&=1\,,\ \ p_1(x)=x\,, \cr
 p_i(x)&= x p_{i-1}(x)- p_{i-2}(x)\ \ \ {\rm for}\ 2\leq i\leq K\,,
  \cr} \eqno(12)$$
The fusion rules of $\phi_1$ read
$$ \eqalign{\phi_1 \times \phi_0 &=\phi_1\,,\ \
\phi_1 \times \phi_K = \phi_{K-1}+\phi_K\,, \cr
\phi_1 \times \phi_i &= \phi_{i-1}+\phi_{i+1}\ \ \ {\rm for}\
i=1,\ldots,K-1\,.} \eqno(13)$$
In the fusion graph $\g_1=A_{q-1}/\Z_2$ (see fig.\ 2)
node $j$ (counting starts
at the free end) represents $\phi_j\,, j=0,\ldots,K\,$. For later
convenience, we would like to order the nodes according to their
label $l(j) = K-j$ so that the connectivity matrix is
$$  N'_1  =  \pmatrix{1 & 1 & 0 & 0 &\cdots  \cr
1 & 0 & 1 & 0 &\ddots \cr 0 & 1 & 0 & 1 &\ddots \cr
0 & 0 & 1 & 0 &\ddots \cr
\vdots & \ddots & \ddots & \ddots &\ddots  \cr}\,. \eqno(14) $$
In the same ordering (indicated by $'\,$) the fusion matrices
of the other fields are $N'_i = p_i(N'_1)\,$. Summarizing, the
proposition is proved if the following lemma holds true:
\mn  {\bf Lemma:} $\sigma_K = p_K(N'_1)\,.$    \mn
It is well known that $N'_1$ has non-degenerate eigenvalues
$\tau^{(m)}=2 cos({m\pi\over q})$, $1 \leq m \leq q-1\,, m$ odd, with
eigenvectors $x^{(m)}$     \def\taum{(\tau^{(m)})}
$$   x^{(m)} = \pmatrix{  p_K(\tau^{(m)}) \cr p_{K-1}(\tau^{(m)}) \cr
  \vdots \cr p_0\taum \cr } \eqno(15) $$
Note that $p_{K+1}\taum =p_K\taum\,$.
The $r$'th component of $\sigma_K x^{(m)}$ is given by
$$ \sum_{l=0}^{K+1-r} p_{K-l}\taum\,. $$  We claim that
$$ \sum_{l=0}^s p_{K-l}\taum = p_K\taum p_s \taum \eqno(16)$$  for
$s=0,\ldots,K\,$, which implies $\sigma_K x^{(m)}=p_K\taum x^{(m)}$
for all $m$, and shows the lemma.         \sn
(16) is proved by induction: Clearly it holds for $s=0,1$. Assume it to
be true up to $s$. Then we have $p_{s+1}p_K=p_1p_sp_K-p_{s-1}p_K$,
and  $$\eqalign{p_1\taum p_K\taum p_s\taum
&= \sum_{l=0}^s p_{K+1-l}\taum + \sum_{l=0}^s p_{K-1-l}\taum  \cr
&= \sum_{l=0}^{s-1} p_{K-l}\taum + p_{K+1}\taum  +
\sum_{l=1}^{s+1} p_{K-l}\taum\,. \cr} $$
Applying the hypothesis for $s-1$ to the first sum in the last line,
and using $p_{K+1}\taum=p_K\taum$, one obtains the statement for $s+1$.
\bn\bn
{\bf 3.}\phantom{xx} In conclusion, the Virasoro minimal models with
central charge $c(2,q)$ have the nice property that their fusion rules
not only encode the modular transformation properties of the
irreducible characters (via Verlinde's formula), but also the whole
character itself, up to a factor involving the central charge.
Put differently: here not only the singular vectors determine
the fusion rules \q1, but also the reverse is true.
Of course, though we do not have any good explanation for this effect,
it would be desirable to have this feature in greater generality,
but an easy extension of the results in the previous section
seems not to be at hand. In a sense, however, the special models
considered above are "basic", because
for each minimal model with $c(p',q')$ there
is a $c(2,q)$ model (with $q=(p'-1)(q'-1)+1\,$, \ie the same number
of primary fields) such that the irreducible representations of
the former are embedded into those of the latter realized as
path spaces as above (not surjectively, however). This can be shown
in the same manner as in \q{8}; note that only the conformal
dimension of the nullfield generating the "annihiliting ideal"
\q{8} has to be used, so perhaps a more thorough study of the
the structure of the relevant normal-ordered products (see
\q{11} for a precise definition) could provide the missing
information in order to construct graphs providing a basis (not only
a spanning system) of the Virasoro representations in question.
\sn
The role of $h_{min}$ in this context can be understood as follows:
$h_{min}$ is the minimal dimension not only of those in
the conformal grid of the $c(2,q)$ theory, but amongst all the
dimensions occuring in minimal models $c(p',q')$ sharing the
same number of primary fields. The asymptotic behaviour of
Virasoro characters suggests to regard the module ${\cal L}_{min}$
as the "largest" occuring in a theory. From this point of view,
it is not so surprising that -- if any fusion graph -- the one of
the minimal dimension field provides the suitable path space to
realize the irreducible representations associated to a given model.
\mn
We also have to comment shortly on the relation between the Virasoro
path realization obtained here and those in \q{3,4}. The ABF and FB
models not only exhibit (multi-)critical points corresponding to
the conformal models of the minimal series (in particular, of the
$c(2,q)$ models), but also allow for an
identification of the Virasoro highest weight modules of the latter
with path spaces occuring naturally in the lattice models (see also
\q{12}). There, the relevant graphs are (unlabeled) $A_{q-1}$
Dynkin diagrams, but the price to be paid for this simplification
consists in the definition of the energy contribution of the
paths (\ie of the $L_0$-operator), which is considerably more involved.
E.g., the module a certain paths belongs to cannot be determined
just from the starting node, but also depends on its infinite
"oscillating tail" (the specific ground state). For the case of
$c(2,q)$ the situation simplifies, as their is only one allowed
ground state \q{3,12}, but except for $c(2,5)$ this path
representation is still very different from the one obtained here.
\sn
Nevertheless, the path algebras \q{13,14} $\a(A_{q-1})$ and
$\a(\g_K)$ associated to the graphs are indeed isomorphic.
This is not obvious from their Bratteli diagrams \q{14},
but can be seen fairly easily from $K$-theoretic arguments: With a
suitable numeration of the nodes we have $\a(A_{q-1},i)=\a(\g_1,i)\,$,
on the other hand $\g_1$ and $\g_K$ share the same Perron-Frobenius
eigenvector $x^{(1)}$, see (15); but this implies that the (scaled)
ordered $K_0$-groups \q{15} coincide: For $s=1,K$ we have
$K_0(\a(\g_s,i))\cong \Z^{K+1}$  with positive cone
$$K_0(\a(\g_s,i))_+\cong \lbrace z\in\Z^{K+1}|\langle z,x^{(1)}\rangle>0
\ {\rm or}\ z=0\rbrace  $$
and order unit $e_i$, the $i\,$th standard basis vector.
Finally, recall that the scaled ordered $K_0$-group is a
complete isomorphism invariant for AF algebras \q{15}.
\sn  This fact suggests that there is indeed a connection between the
two path realizations, so maybe it would be amusing to define an
RSOS model based on the fusion graphs considered here and to see if
it can be mapped to the corresponding FB model.
\mn
The last paragraph leads us to another reason why we think path
realizations of Virasoro highest weight modules are important:
Their existence implies the possibility of
an AF algebra representation of the Virasoro algebra itself.
Having established a grading-preserving isomorphism between the $\l_j$
of some model and path spaces on a graph, one may conclude that
in principle there is a path algebra representation of the Virasoro
algebra (with given $c$) itself, at least in the sense that
the Virasoro generators can be expressed as limits (infinite sums)
of elements in this AF algebra -- which is generated by
Temperley-Lieb-Jones (TLJ) projections and
possibly some finite dimensionsal matrix algebra. Although
finding the explicit relation between Virasoro algebra and
TLJ generators \q{16} is still a very difficult task,
its mere existence already has important implications,
possibly up to new approaches to a classification of CFT using the
framework of Algebraic QFT \q{17} together with invariants
of $C^*$-algebras, see \q{18} for first attempts in this direction.
\bn  \bn
We would like to thank W.\ Nahm, W.\ Eholzer, M.\ Roesgen and R.\
Varnhagen for various discussions on related subjects. \hfil\break
\noindent A.R.\ is supported by the Max-Planck-Institut f\"ur
Mathematik, Bonn.
\bn\vfill \eject  \leftline{{\bf References }} \bn
\halign{\hfil $ # $\phantom{xx}& \rm # \hfill  \cr
\q{1} &A.A.\ Belavin, A.M.\ Polyakov, A.B.\ Zamolodchikov,
       \np {\bf 241} (1984) 333; \cr
      &D.\ Friedan, Z.\ Qiu, S.H.\ Shenker, Phys.\ Rev.\ Lett.\
       {\bf 52} (1984) 1575 \crns
\q{2} &R.J.\ Baxter, {\sl Exactly solved models in statistical
       mechanics}, New York 1982 \crns
\q{3}&G.E.\ Andrews, R.J.\ Baxter, P.J.\ Forrester,
      J.\ Stat.\ Phys.\ {\bf 35} (1984) 193; \cr
      &P.J.\ Forrester, R.J.\ Baxter,
       J.\ Stat.\ Phys.\ {\bf 38} (1985) 435 \crns
\q{4} &E.\ Date, M.\ Jimbo, A.\ Kuniba, T.\ Miwa, M.\ Okado,
        \np {\bf 290} (1987) 231; \cr
      &E.\ Date, M.\ Jimbo, T.\ Miwa, M.\ Okado, Proc.\ Symp.\ Pure
       Math.\ {\bf49} (1989) 295  \crns
\q{5}  &V.\ Pasquier, J.\ Phys.\ A {\bf20} (1987) 5707, \np
        {\bf285} (1987) 162,  \cr
       &{\bf295} (1988) 491, \CMP {\bf118} (1988) 355 \crns
\q{6}  &A.\ Cappelli, C.\ Itzykson, J.-B.\ Zuber, \CMP
        {\bf113} (1987) 1; \cr
       &D.\ Gepner, \np {\bf287} (1987) 111 \crns
\q{7}  &H.\ Saleur, in {\sl Knots, Topology and QFT}, Proceedings of
        the \cr
       &John Hopkins Workshop, Florence 1989, ed.\ L.\ Lusanna \crns
\q{8}  &B.L.\ Feigin, T.\ Nakanishi, H.\ Ooguri, Int.\ J.\ Mod.\ Phys.\
        A {\bf7} Suppl.\ 1A, (1992) 217  \crns
\q{9}  &V.G.\ Kac, M.\ Wakimoto, Proc.\ Natl.\ Acad.\ Sci.\ USA
        {\bf 85} (1988) 4956 \crns
\q{10} &M.\ Caselle, G.\ Ponzano, F.\ Ravanini, \pl {\bf251} (1990)
        260; \cr
       &{\sl Towards a classification of fusion rule algebras in
        rational conformal} \cr
       &{\sl field theories}, Saclay preprint SPht/91-174  \crns
\q{11} &R.\ Blumenhagen, M.\ Flohr, A.\ Kliem, W.\ Nahm, A.\ Recknagel,
        R.\ Varnhagen, \cr &\np {\bf 361} (1991) 255 \crns
\q{12} &H.A.\ Riggs, \np {\bf 326} (1989) 673 \crns
\q{13} &A.\ Ocneanu, in {\sl Operator Algebras and Applications II},
       London Math.\ Soc. \cr &vol.\ 135, Cambridge 1988;  \cr
       &{\sl Quantum Symmetry, Differential Geometry of Finite
        Graphs and Classification} \cr  &{\sl of Subfactors},
       University of Tokyo Seminary Notes 45, July 1990; \crns
\q{14} &F.M.\ Goodman, P.\ de la Harpe, V.F.R.\ Jones, {\sl Coxeter
        Graphs and Towers} \cr
       &{\sl of Algebras}, MSRI Publ.\ 14, New York 1989 \crns
\q{15} &B.\ Blackadar, {\sl K-Theory for Operator Algebras}, MSRI
        Publ.\ 5, New York 1986  \crns
\q{16} &M.\ R\"osgen, R.\ Varnhagen, in preparation \crns
\q{17}&S.\ Doplicher, R.\ Haag, J.E.\ Roberts, \CMP {\bf23}
       (1971) 199, \cr  &{\bf35} (1974) 49; \cr
       &K.\ Fredenhagen, K.H.\ Rehren, B. Schroer, \CMP
       {\bf 125} (1989) 201 \crns
\q{18} &A.\ Recknagel, {\sl Fusion rules from algebraic K-theory},
        Bonn University preprint \cr
       &Bonn-HE-92-06, to be published in \intj   \crns }
\bn\bn\bn  \leftline{\bf Figure captions}  \bn
\item{Figure 1:} Fusion graph of the $h_{min}$ primary field in
the $c(2,5),\ c(2,7)$, and $c(2,9)$ minimal model. The labels
are as in the proposition.   \mn
\item{Figure 2:} Fusion graph of the "generating field"
$\phi_{(q-2,1)}$ for $c(2,q),\ q=2K+3\,$; labeling as above.
\mn   \vfill\eject\end